# Nanodiamond-Enabled Torsion Microscopy Uncovers Multidimensional Cell-Matrix Mechanical Interactions

*Yong Hou[1], Lingzhi Wang[1], Zheng Hao[2], Fuqiang Sun[2], Yutong Wu[3], Luyao Zhang[1], Linjie Ma[1], Wenyan Xie[4], Xinhao Hu[1], Qiang Wei[5], Cheng-han Yu[3*], Yuan Lin[2*], Zhiqin Chu[1,3,6*]*

1. Department of Electrical and Electronic Engineering, The University of Hong Kong, Hong Kong.
2. Department of Mechanical Engineering, The University of Hong Kong, Pok Fu Lam, Hong Kong.
3. School of Biomedical Sciences, The University of Hong Kong, Pok Fu Lam, Hong Kong.
4. Department of Biomedical Engineering, Hong Kong Polytechnic University, Kowloon, Hong Kong.
5. College of Polymer Science and Engineering, State Key Laboratory of Polymer Materials and Engineering, Sichuan University, Chengdu, Sichuan, 610065, China.
6. School of Biomedical Engineering, The University of Hong Kong, Pok Fu Lam, Hong Kong.

\* Corresponding authors:
Prof. Dr. Cheng-han Yu, Email: chyu1@hku.hk
Prof. Dr. Yuan Lin, Email: ylin@hku.hk
Prof. Dr. Zhiqin Chu, Email: zqchu@eee.hku.hk

**Abstract**

Traditional cellular force-sensing techniques, such as traction force microscopy (TFM), are predominantly limited to measuring linear tractions, overlooking and technically unable to capture the nanoscale torsional forces that are critical in cell-matrix interactions. Here, we introduce a nanodiamond-enabled torsion microscopy (DTM) that integrates nitrogen-vacancy (NV) centers as orientation markers with micropillar arrays to decouple and quantify nanoscale rotational and translational motions induced by cells. This approach achieves high precision (~1.47° rotational accuracy and ~$3.13 \times 10^{-15}$ N·m torque sensitivity), enabling reconstruction of cellular torsional force fields and twisting energy distributions previously underestimated. Our findings reveal the widespread presence of torsional forces in cell-matrix interactions, introducing "cellular mechanical modes" where different adhesion patterns dictate the balance between traction- and torque-mediated mechanical energy transferred to the substrate. Notably, in immune cells like macrophages that generally exert low linear tractions, torque overwhelmingly dominates traction, highlighting a unique mechanical output for specific cellular functions. By uncovering these differential modes, DTM provides a versatile tool to advance biomechanical investigations, with potential applications in disease diagnostics and therapeutics.

**Introduction**

Cells probe and remodel their extracellular matrix (ECM) through multidimensional physical forces, guiding essential processes like migration, immune response, and development (*1–3*). These forces, transmitted through adhesion structures such as focal adhesions and podosomes, are conventionally categorized into two well-studied types: in-plane shear tractions that drive "push-pull" motility, and normal forces that compress or stretch the substrate vertically. Emerging evidence suggests the existence of a third, more elusive interaction – rotational moment between the cell and the surrounding matrix. For instance, podosome rings can generate significant torsions on the substrate, distinct from linear pulling forces of focal adhesions (*4*). Similarly, substantial out-of-plane, lever-like rotational moments have also been found on focal adhesions (*5*). These rotational forces actively remodel the ECM and modulate mechano-signaling, influencing critical processes such as cell migration(*6*), immune reaction (*7–9*) and embryonic organ development (*10, 11*). However, the spatiotemporal dynamics of rotational moments and their precise linkage to adhesion structures remain poorly understood.

This gap stems primarily from the limitations of current force-sensing technologies. Specifically, conventional techniques like traction force microscopy (TFM) and molecular tension microscopy (MTM) rely on measuring the linear displacement of the fluorescent beads or the one-dimensional extension of 'spring-like' ligands (e.g., polyethylene glycol, DNA hairpins, or proteins) (*12–15*). While these approaches are powerful for measuring linear forces they fall short in capturing rotational motions, such as twist of the ECM or cytoskeleton/membrane proteins, due to three key reasons: 1) bead displacements or ligand extensions only provide 1D or 2D translational readouts; 2) torsional signals are often overlapped with those induced by linear tractions; 3) lack specialized markers capable of reporting precise angular changes.

Nitrogen-vacancy (NV) centers - defects in the diamond lattice, have emerged as one of the most promising platforms for quantum biosensing (*16, 17*). Their unique spin-dependent fluorescent properties and outstanding sensitivity and spatial resolution have driven a surge of interest in various applications, including monitoring physiological signals such as intracellular temperature (*18*), free radicals (*19, 20*), and pH levels (*21*). In addition, the dipole-like features of NV centers exhibit intriguing characteristics when subjected to polarized light, a phenomenon known as optical polarization selective excitation. This property allows for the determination of the axis orientation of individual NV centers within bulk diamond (*22*). Recently, we harnessed this unique polarization property of NV centers in nanodiamonds as fluorescent orientation markers and developed a linear polarization modulation (LPM) method to monitor the in-plane rotational motions of these nanodiamonds with high precision, achieving an accuracy of 0.5-3°. In this study, we present a novel approach that ingeniously integrates nanodiamonds with classical micropillar arrays to create a nanodiamond-enabled torsion microscopy. This innovative design enables precise quantification of cellular torsional forces at the nanoscale, addressing limitations of traditional measurement techniques. It offers new mechanical insights into cell-matrix interactions, with potential to advance mechanobiology, immunology, and pathological mechanics, and possible future applications in disease diagnosis and therapy.

**Results**

**System design and working principle of DTM**

Cellular forces are typically coupled and act collectively on the substrate, posing a significant challenge in separating and quantifying torsional forces from linear tractions. To address the challenge, we developed a DTM that combines a fluorescence polarization confocal microscope and a hybrid nanodiamond sensor array as the optical detector and force transducer, respectively (**Fig. 1a**). In the optical setup, a linearly polarized 532 nm laser passes through a polarization modulator (PM) to vary the excitation polarization angle, enabling the detection of in-plane rotational changes in the NV centers. These NV centers are embedded within nanodiamonds (NDs) that are conjugated to the top of PDMS micropillars. Each pillar acts as an independent sensing unit capable of detecting multi-degree-of-freedom motions, including translation and rotation (Fig. 1b). The NV center, formed by a substitutional nitrogen atom adjacent to a vacancy in the diamond lattice, serves as a fluorescent orientation marker. Its dipole moment, perpendicular to the NV axis, exhibits optical polarization selective excitation properties, whereby the fluorescence intensity varies with the angle between the laser polarization and the NV axis (Fig. 1c).

The core innovation of the DTM lies in decoupling cellular torsional moments from linear tractions through the pillar's deflection and the NV's optical polarization sensitivity to in-plane rotations. Unlike contiguous substrates where force applied at one point could influence deformation in nearby regions, the discrete pillar array ensures independence between sensing units, minimizing force interference and enabling localized measurements at individual adhesion sites (typically 1-5 μm). When a cell applies an off-center force on a pillar, it induces both translational bending (reporting traction via pillar deflection) and rotational torsion (reporting torque via ND rotation detected through NV signals). Specifically, the optical polarization anisotropy depends on orientation (*22*), with the effective excitation rate of NV centers ($I_{effect}$) depends on the relative angle ($\alpha - \beta$) between laser polarization and the projection of each NV axis onto the sample surface (*23*). The relationship can be described as:

$$I_{effect} = I_{actual} \left[\frac{8}{9}(1 - cos^2(\alpha - \beta)sin^2(\gamma)) + \frac{1}{9}\right] \quad (1)$$

where the $I_{actual}$ indicate the actual excitation rate. The angle $\alpha, \beta,$ and $\gamma$ are shown in Fig. 1c. Thus, the orientations of the projected NV centers in-plane can be distinguished optically, allowing quantitative detection of rotational motions of nanodiamond through polarized scanning confocal imaging (Fig. 1d). This integration positions DTM as a promising platform for decoupling multidimensional cellular forces.

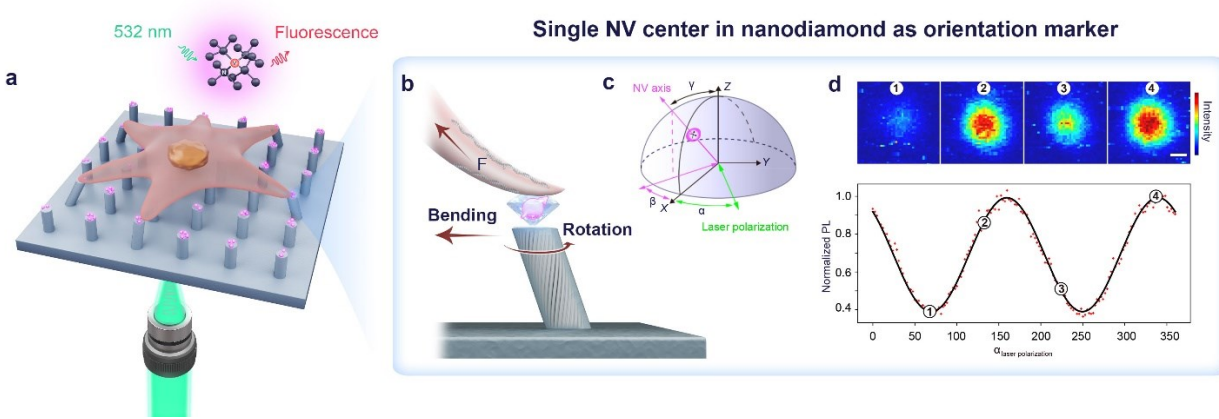

**Figure 1. Conceptual design of nanodiamond-enabled torsion microscopy (DTM).** a) Diagram showing DTM setup. A linearly polarized laser (532 nm) passes through a polarization modulator (PM) to modulate the polarization light direction within the plane. The NV center (violet arrow) in diamond as the orientation marker on the top of pillar. Insert shows the NV center is formed by a substitutional nitrogen atom and an adjacent vacancy in the diamond lattice. b) The structure and working principle of the ND@pillar-based torsion probe. c) Schematic illustration showing the NV axis orientation (violet thick solid arrow) with corresponding projection (violet solid arrow) and laser polarization direction (green arrow). Black thin solid arrows are two transition dipoles perpendicular to each other, lying in the plane perpendicular to the NV axis. d) A typical LPM curve of the signal NV center in diamond. Upper panel: Fluorescence images of a single NV center

under laser excitation with different polarization angles. Lower panel: Corresponding LPM curve with experimental data points and fitted model. Scale bar: 250 nm.

**Characterization and validation of the DTM System**

To validate the torque sensing capability and explore how pillar physical properties influence performance, we first simulated PDMS micropillar deformation using COMSOL Multiphysics® software under the Solid Mechanics interface. For circular pillars (diameter D = 2 µm, Young's modulus E = 1.4 MPa), a 50 nN tangential force applied at varying offsets from the center induces significant rotation alongside translation (**Fig. 2a**). The rotational angle scales linearly with the loading position offset and inversely with the aspect ratio (L/D), reaching up to ~20° for edge loading at L/D = 3 (Fig. 2b and Fig. S1). To experimentally validate these simulation predictions and directly visualize the torsional effect, we employed a PDMS pillar array with square-shaped pillars. The distinct edges of the square geometry enable unambiguous identification of rotational displacements via scanning electron microscopy (SEM), which is challenging with circular pillars in conventional pillar-based traction force microscopy. As shown in Fig. 2c, a NIH-3T3 cell fully spread and randomly anchored to the top surface of square pillars, inducing prevalent rotational motions with maximum angles approaching approximately 9.1° (Fig. 2c and Fig. S2). Our observations revealed that the anchoring positions of filopodia and the elongated cell body (i.e., force loading points) rarely aligned with the geometric center of the pillar top surface (Fig. S2) and therefore led to torque generation. Similar misalignments were also commonly observed on circular pillar arrays (Fig. S3), demonstrating the widespread yet often overlooked presence of torsional forces in cell-pillar interactions. Together, these findings confirm that cells readily induce both bending and twisting in pillars, validating our theoretical framework and positioning DTM as a robust platform for decoupling multidimensional cellular forces in mechanobiology studies.

We next characterized the sensors and DTM's performance. The ND samples used in this study were pre-treated via a salt-assisted air oxidation (SAAO) method (*24*), resulting in a uniform size distribution of 177 ± 47 nm and a negatively charged surface with zeta potential of –28 mV (Fig. S4). We successfully conjugated these NDs to the tops of PDMS micropillars via electrostatic adsorption(*25*), achieving a stable and controllable density of ∽ 8 NDs per pillar used in subsequent experiments (Fig. 2d and Fig. S5). To characterize the rotational sensing performance of the ND@pilar sensors, a half-wave plate was rotated in the excitation path to generate linearly polarized laser light at controlled angles. The resulting optical polarization dependence of a typical ND@pilar is shown for four angles in Fig. S1d. By varying excitation angle ($\alpha$) and measuring total fluorescence, we determined the projection of the NV's orientation onto the plane by defining the $\beta$-value corresponding to the minimum fluorescence intensity in the LPM fitting curve. The nanodiamond's rotational angle ($\Delta\beta$) was then calculated from the shift in $\beta$ between different force states. The accuracy of the orientation measurement is critically dependent on the contrast value determined by $\gamma$. Higher contrast in the LPM signals enables more precise determination of the NV center's orientation(*23*). Impressively, the SAAO-treated NDs exhibited a high yield of single NV centers (~30%) and a high average LPM contrast of 30.7% (Fig. S6), which translated to an excellent measurement precision of ~1.05° (Fig. 2e). The rotation measurement error was determined at 1.47 ± 1.15° (Fig. 2f). Consequently, the overall precision for measuring cellular torque in the current setup is $3.13\times10^{-15}$ N·m (Fig. 2g). Notably, the torque sensitivity can be further enhanced by using mechanically more compliant pillars. Increasing the pillar aspect ratio (L/D) or decreasing the pillar stiffness dramatically improve measurement sensitivity (Fig. 2g and Fig. S7).

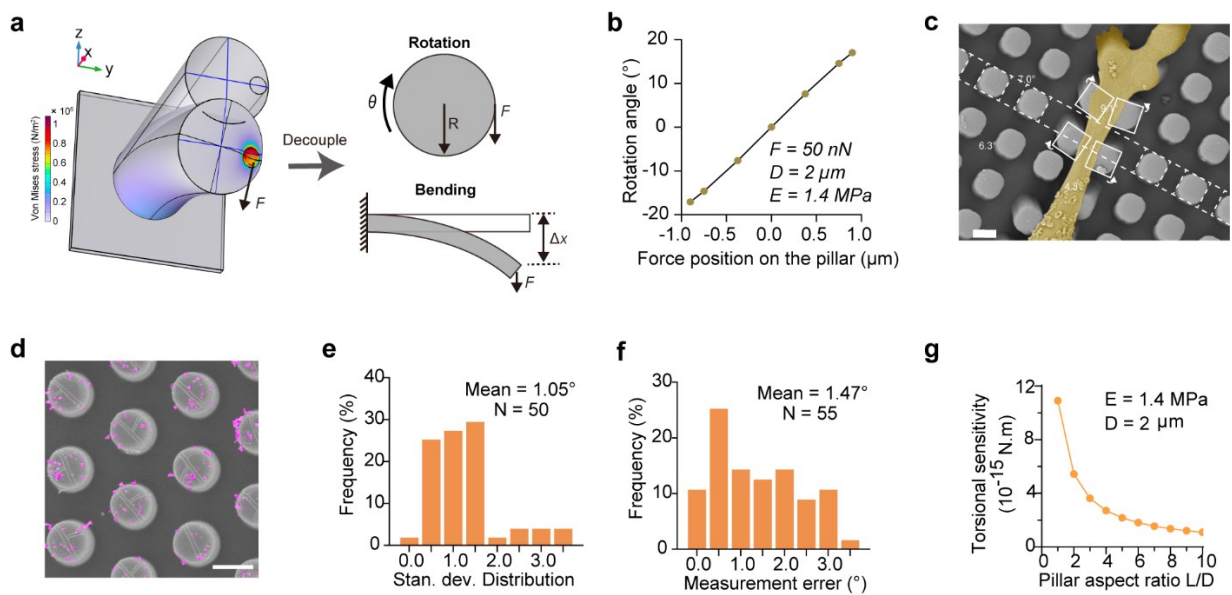

**Figure 2. Validating and evaluating the performance of DTM.** a) COMSOL simulation demonstrates the existence of torque on the pillar when the cell traction force is not located at the geometrical centerline of the pillar. b) Modeling the pillar rotational angels as the function of force loading positions. c) SEM image of cell spreading on the square-shaped PDMS pillar array (The cell is labelled with golden color). The solid rectangular box represents the outline of the target micropillar, and the dashed line represents the reference line for direction. Scale bar: 2 μm. d) SEM image representing the successful modification of NDs on the top surface of PDMS pillars. NDs are labeled with violet color. Scale bar: 2 μm e) Histogram of the standard deviations of orientation determination by the LPM fitting in different NDs. LPM tests for each diamond were duplicated 4 times. f) Histogram representing the measurement error in rotational motion measurements. g)

Torque measurement precision of DTM as the function of aspect ratio (L/D). *D*: Pillar diameter, *E*: Young's modulus.

**Quantification of the cellular torque and twisting energy by DTM**

To demonstrate the capabilities of DTM, we measured both the rotational and translational motions of ND-fibronectin-coated pillars induced by NIH-3T3 cells. The null-force condition (force-relaxation state) was subsequently obtained after removing the cell from the pillars. The pillar displacement was obtained by a conventional pillar recognition program used in TFM (**Fig. 3a**). Using DTM, we determined the orientation of the NV projection through LPM curve fitting (See fitting process in Methods) and pillar rotations induced by the cell were quantitatively measured (Fig. 3b, 3c and 3e). As expected, larger pillar rotations (6.6 ± 4.4°) were observed at cell boundaries compared to areas beneath the cell body (2.9 ± 2.1°) or non-cell adhesion regions (1.29 ± 1.26°). Notably, three pillars at the cell's leading edges exhibited pronounced rotations ranging from 7° to 14.9°, coinciding with areas of increased cellular traction confirmed by TFM (Fig. 3b, 3c, and Fig. S8). A significant positive correlation between pillar rotational and translational motions was noted at the cell boundary, suggesting that translational forces may contribute to torque generation in NIH-3T3 cells, particularly at the cell periphery where focal adhesions are prominent (Fig. 3f).

Beside the force magnitude, the position of loading determining the arm of force can also affect pillar rotation (Fig. 2a and 2b). Indeed, the off-center force loading points on the pillar, revealed by filamentous actin (F-actin) staining can be used to estimate the directions of pillar rotation. For example, both the downward contraction force on the left edge of pillar 1 and the upward contraction force on the right edge of pillar 2 led to an anticlockwise rotation of these pillars, and the corresponding ND-based fluorescence signals were also detected (Fig. S8). This intuitive inference was confirmed by our LPM results, showcasing the reliability of our technology.

One of the unique advantages of DTM is to quantify the torque and twisting energy generated by cells on each pillar. We found that the order of magnitude of cellular torque and twisting energy applied on pillar were estimated to be around $10^{-14}$ N·m and $10^{-16}$ - $10^{-15}$ J, respectively (See calculation process in Methods). Remarkably, this twisting energy $W_{twisting}$ was comparable to the bending energy ($W_{bending}$) (Fig. 3g) stored in the pillar, highlighting the previously underappreciated significance of torsional forces in cell-matrix interactions. Interestingly, the twisting energy predominated at the cell boundary, with a $W_{twisting}/W_{bending}$ ratio of ~1.2 which was much larger than that of ~0.5 beneath the cell body (Fig. 3h). Immunofluorescence staining of vinculin revealed the enrichment of focal adhesions at the cell boundary exhibited asymmetrical state, showing a curved structure located at the edges of the pillars (Fig. 3d and Fig. S9). This spatial arrangement of adhesion results in a long "arm of force" exerted on the pillar, and therefore a higher percentage of twisting energy stored.

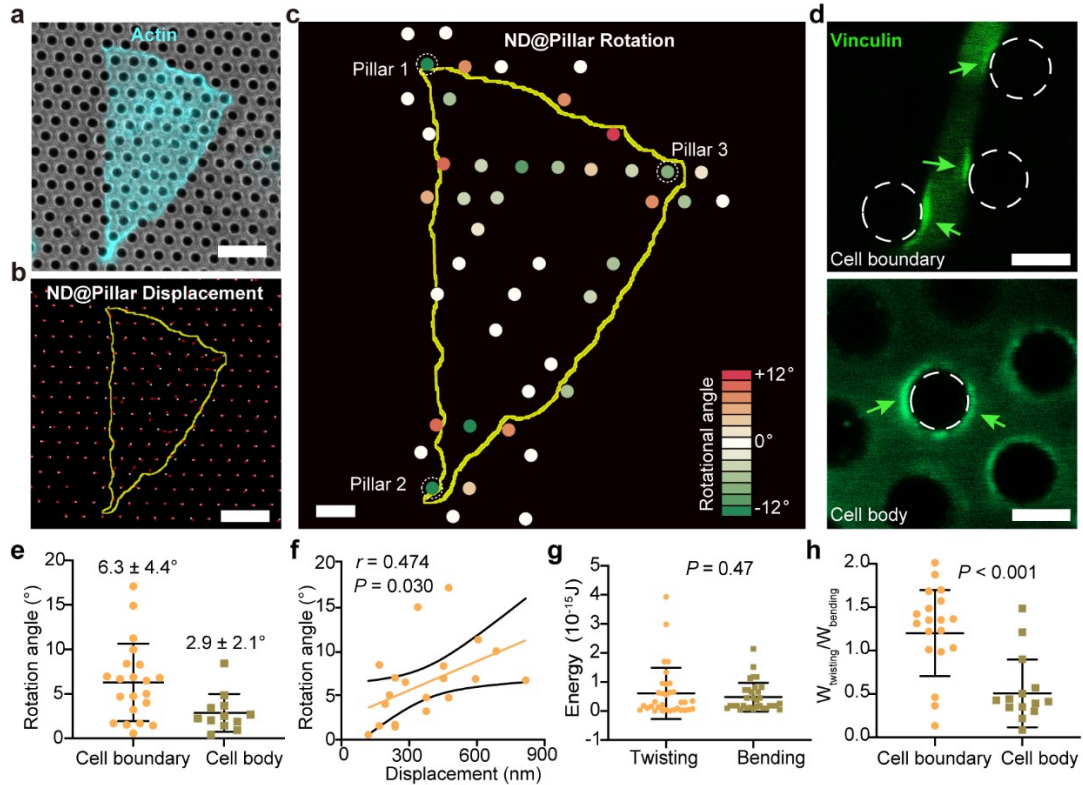

**Figure 3. Revealing the underestimated rotational motions of micropillars interacting with NIH-3T3 cells** a) The fluorescence image of NIH-3T3 cell spreading on pillar array. Turquoise (actin). Scale bar: 10 μm. b) Displacement mapping of pillar array in NIH-3T3 cell force field. Scale bar: 10 μm. c) Color-coded rotational map of pillars in cell force field. Scale bar: 5 μm. d) Representative fluorescence images show the vinculin morphology (green) at cell boundary and cell body. The white dashed circles represent the pillar positions. Scale bar: 2 μm. e) Rotation angle of the pillars located at the cell boundary (N=21) and underneath the cell body (N=12), respectively. f) Correlation analysis revealing the relation between rotational motions and translational motions of pillars at the cell boundary (N=21). Pearson's correlation coefficient, $r$, between pillar translational and rotational motions. $P < 0.05$ indicates statistically significant correlation of a linear regression. g) The twisting energy and bending energy stored in the targeted pillars (N = 33). h) The distinct ratio of twisting energy and bending energy stored in pillars located at the cell boundary (N=21) and underneath the cell body (N=12), respectively.

**DTM captures the torsional forces dominating macrophage-substrate interactions**

Building on our findings that NIH-3T3 cells generate substantial torsional forces at the cell periphery via focal adhesions, we next sought to investigate whether this phenomenon is cell-type specific. We applied DTM to macrophages, which form highly dynamic podosomes, exert lower translational forces than fibroblasts (*26*), and have been reported to produce torsional tractions during podosome reorganization (*4*). These features make macrophages an ideal model for challenging DTM's sensitivity in low-force regimes and ability in capturing faint, spatially distinct mechanical signals.

Consistent with prior reports, macrophages exerted weak translational forces, with most pillar displacements below 300 nm (under the optical diffraction limit) and thus were hardly detectable by conventional TFM with spatial resolutions of micrometers (*27*) (**Fig. 4a,** b and f). In contrast, DTM revealed substantial rotational deformations, with maximum angles reaching 15° (Fig. 4c and d), underscoring DTM's high torque sensitivity (~1.47° precision). Unlike the peripheral distribution of torque in 3T3 cells (Fig. 3e), macrophage-induced rotations were spatially uniform, showing no significant difference between cell edges (4.9 ± 3°) and the cell body (5.1 ± 4.8°; Fig. 4d). Unlike NIH-3T3 cells, macrophages assemble prominent "ring-like adhesions" with asymmetrical enrichment of vinculin around the pillars across the entire ventral surface (Fig. 4a, S10). In addition, the translational and rotational movement of pillars were not positively correlated (Fig. 4f), suggesting that the off-center horizontal tractions may not dominate the rotational force but may involve intrinsic twisting mechanisms at macrophage adhesion sites.

When comparing the bending and twisting energy exerted by macrophage, the $W_{twisting}/W_{bending}$ ratio was ~4.4 at the cell edge and ~2.2 under the cell body, with no significant regional difference ($P = 0.255$, Fig. 4g). Intriguingly, twisting energy overwhelmingly dominated the bending energy at all pillars under macrophages (Fig. 4e), far exceeding the ratios seen in NIH-3T3 cells (Fig. 3h). These measurements highlight the unique advantage of DTM in unveiling additional information of torsion, in spite of non-detectable traction force. Torque, rather than traction, is the primary mode of force transduction at the adhesion interface of pillar and macrophage.

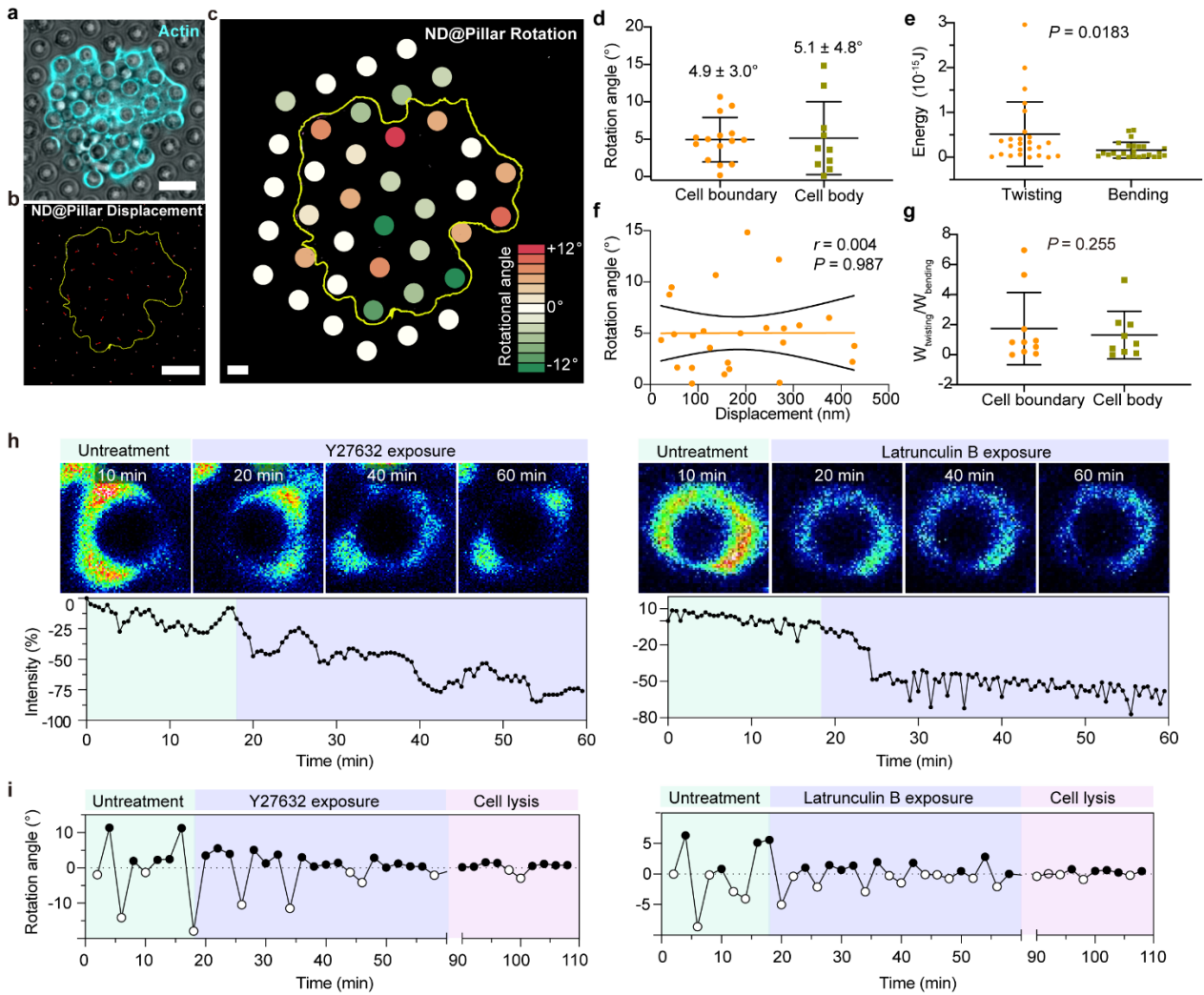

**Figure 4. Uncovering the torque-dominant forces exerted by macrophages.** a) Fluorescence image of macrophage spreading on pillar array. Turquoise (actin). Scale bar: 5 μm. b) Displacement mapping of pillar array. Scale bar: 10 μm. c) Color-coded rotational map of pillars in cell force field. Scale bar: 5 μm. d) The rotation angle of the pillars located at the cell boundary (N=15) and underneath the cell body (N=10), respectively. e) The twisting energy and bending energy stored in the targeted pillars (N=25). f) Correlation analysis reveals the relation between rotational motions and translational motions of pillars at the cell boundary and underneath the cell body (N=25). Pearson's correlation coefficient, $r$, between pillar translational and rotational motions. $p < 0.05$ indicates statistically significant correlation of a linear regression. g) The ratio of twisting energy and bending energy stored in pillars located at the cell boundary (N=14) and underneath the cell body (N=9), respectively. h) Live fluorescent imaging of ring-like adhesion complex of macrophage on micropillar sensors treated with different inhibitors. Fluorescent images of THP-1 macrophages expressing lifeact-EBFP2 (a F-actin marker) are taken at 30s intervals over a 60-

minute period. The line chart indicates the fluorescence intensity dynamics of the "ring-like adhesion complex" under different chemical treatments. i) Representative profiles of ND rotational angles in macrophages cultured on micropillar substrates with different pharmacological treatments. Measurements are conducted at 2-minute intervals over 110 min, with a 30-minute pause for cell lysis treatment. Data represents changes in rotational angle calculated as differences between consecutive time points.

We next explore whether DTM could capture the dynamics of torsional activity in live cells. Macrophages were seeded on the ND-fibronectin-coated pillars. The cytoskeletal reorganization and pillar motions were seamlessly recorded by confocal microscopy and DTM, respectively. In control condition, pillars exhibited continual oscillations in both translation (0-200 nm) and maximum rotation (~10°), reflecting the highly dynamic nature of the ring-like adhesion complexes (Fig. S11, Video 1). To further examine the functional role of actomyosin in the torsional force development, ROCK inhibitor Y27632 (20 μM; block myosin-II activity) or Latrunculin B (10 μM; block F-actin polymerization) was introduced. Upon Y27632 treatment, actin fluorescence intensity at adhesions decreased gradually over 40 minutes, accompanied by a progressive decay in the maximum rotational oscillation amplitude—from 17.9° before treatment to ~10° within the first 20 minutes after treatment, eventually declining to ~2° (near the measurement noise floor observed in lysed cells) by the end of the recording (Fig. 4h, i, Video 2). Latrunculin B induced rapid and dramatic disruption of adhesion dynamics: actin fluorescence intensity dropped sharply by ~50% within the first 10 minutes and then stabilized, while the maximum rotational amplitude decreased abruptly from 8.6° to ~2° over the same period and remained low thereafter (Fig. 4h, i, Video 3). These distinct response patterns highlight that myosin-II mediated contractility plays a pivotal role in driving rotational oscillations, while the assembly of F-actin acts as the structural component of force transmission at the macrophage adhesion.

**Discussion**

Force transmission at the cell-matrix interface is fundamentally multidimensional, yet our understanding of it has often been constrained by limited-dimensional force tools. By integrating NV-center sensors with micropillar arrays, DTM enables quantitative decoupling translational and rotational forces, demonstrating that cells not only can pull or push the substrate but also twist. In addition, the twisting energy, resulted from the torque development is comparable to, and in some cases can dominate the bending energy. This discovery challenges the traction-centric view of cell-matrix interactions and provides new insight into the intricate cell-matrix force transduction.

Our results highlight distinct mechanical modes of interaction between the cell and substrate (micropillars) in a cell-type specific manner. NIH-3T3 cells operate in a "traction-dominated mode". Torsional forces are concentrated at the cell periphery via focal adhesions, resulting in a twisting-to-bending energy ratio of ~1.2 at edges compared to ~0.5 beneath the cell body. The low twisting-to-bending energy ratio implies that torque arises as a byproduct of off-center tractions, and the traction-dominant mode orchestrated by the fibroblast facilitates large-scale matrix remodeling. In contrast, macrophages exhibit a "torque-dominant mode", with spatially uniform torque distribution and twisting energy overwhelmingly dominating bending energy (Fig 4d-g). This uniformity aligns with the random distribution of ring-like adhesion complexes across the macrophage ventral surface (Fig. 4a and Fig. S11). Furthermore, the decoupling of translational and rotational movement of pillar underneath the macrophage implies an intrinsic twisting mechanism at the ring-like adhesion complex. The ROCK inhibitor Y27632, which disrupts myosin-II contractility, causes a gradual decay in rotational oscillations over 40 minutes, suggesting that myosin-II-based tension is critical for sustaining the torsional activity of these complexes. In contrast, the actin depolymerization induced by Latrunculin B elicits rapid mechanical quenching within minutes, directly correlating the disassembly of the actin cytoskeleton with the immediate cessation of torque generation. Our findings align with prior observations in podosomes, where similar inhibitor-based experiments demonstrated that podosome rotation and torsional traction on planar substrates depend on actomyosin contractility and actin dynamics (*4*). These torsional tractions have been shown to deform the local mechanics of the substrate, promoting the stability of the adhesion core, and facilitating ECM degradation. Building on this, we hypothesize that this torque-dominant mode may enable macrophages to perform different modes of mechanical probing in complex environments, preparing for subsequent immune responses like migration, phagocytosis, or localized matrix degradation—potentially an energy-efficient strategy in low-force regimes.

These new discoveries stem from DTM's ability to address core challenges in multi-dimensional force decoupling and rotational signal detection—issues intractable for traditional TFM or MTM. DTM's three key advantages include: (1) analytical decoupling the degree-of-freedom, separating rotation from translation; (2) rotational angle detection, using NV centers as the unique orientation markers for rotation tracking; and (3) high sensitivity, extracting the underlying mechanical information from weak force fields (e.g., in macrophages) where traditional TFM fails. Importantly, DTM's torque sensitivity is not fixed but can be dramatically enhanced through sensor engineering. For instance, by decreasing the diameter to 1 μm while maintaining aspect ratio of 3 and Young's modulus of 1.4 MPa— the torque sensitivity can be enhanced to a level that corresponds to a force of only ~784 pN applied at the edge of the pillar (Fig. S8). This force equivalent is far below the typical detection limit of conventional TFM (typically in the 2-120 nN range) (*28*), makes it highly suitable for measuring weak force fields. Additionally, the size of diamond with single NV center is tunable and the commercially available nanodiamonds can be as small as ~5 nm, which are comparable to various mechanosensitive components in cells, such as integrin (8-12 nm), T cell receptor (7.5 nm), and microtubules (25 nm), (*3*, *29*, *30*) highlighting the potential of torsional force measurement in a single-protein level.

As the first-generation cell torsion microscope, the current DTM is limited by its temporal resolution (15 s/loop for a single LPM test). A widefield-based ultrafast PM now under development promises millisecond-scale temporal resolution, enabling detailed mappings of cellular torsional dynamics soon. Additionally, although pillar-based force sensors are widely used in the field, their special topological structure, unlike traditional planar substrates, may modulate cell adhesion structures and mechanosensing. To minimize interface topology effects, future designs could incorporate smaller-scale pillar arrays (e.g., nanoscale pillar diameter and spacing).

DTM is a powerful tool for non-invasively monitoring multidimensional mechanical signals (translation, rotation, and energy) exerted by the cell. This capability, unattainable with traditional techniques, unveils that torsion, rather than traction is the dominant mechanical signal at the macrophage adhesion. Looking ahead, the diamond-based torsion microscope and the newly revealed force – "torsional force", may provide new insight into the mechano-immunology researches. The capability to decouple torsion and traction establishes a new foundation for future investigations into whether torsional forces influence immune-specific processes, such as phagocytosis, antigen presentation, or extracellular matrix remodeling. Given its simplicity in sensor fabrication and compatibility with fluorescence microscopy, we anticipate that our DTM will serve as a powerful tool in driving mechanobiological advancements.

## Methods

### Optical setup

The experiments were performed on a home-built confocal microscope reported previously.(*23*) Specifically, in excitation optical path, a polarizer (FLP25-VIS-M, LBTEK) is used to ensure the high linear polarization ratio of 532 nm excitation laser. A half-wave plate (HWP20-532BT, LBTEK) is inserted in the electrical rotation stage (PT-GD62, PDV) to change the polarization direction of the excitation laser. With the rotation speed of 12°/s, it takes 15s to measure two periods of the LPM curve. The fluorescence signal is collected in the detection path by single-mode fused fiber optic couplers (TW670R5F1, Thorlabs) with two single photon counting modules (SPCM-AQRH-16-FC, Excelitas Technologies) as detectors. Time tagger 20 (Swabian) is used for measuring the second-order correlation function $g^2(\tau)$. In addition, a transmission light path is coupled into the confocal microscope for bright-field imaging.

**Fabrication of torsional force sensors**

Step 1: prepare the PDMS pillar array.

The PDMS micropillar array was fabricated as previously described using standard soft lithography technique.(*31*) The dimensions of individual pillars were 6 μm in height and 2 μm in diameter. For square-shaped pillar array, the pillars were 6 μm in height and 2 μm in the side length. Specifically, PDMS (mixed at 10:1 with its curing agent, Sylgard 184; Dow Corning) was then poured over the silicon mold (cut into 1 cm * 1 cm slide), vacuumed for 30 minutes and then covered with a cove slides (2.2 cm * 2.2 cm, ∽140 μm in thickness). The obtained samples were then cured at 80 °C overnight to reach a Young's modulus of 1.4 ± 0.5 MPa and peeled off very carefully. Pillar arrays were hexagonal, with center-to-center spacing set to twice the pillar diameter to maintain constant areal density. Considering the working distance of the objective in our optical setup is only 280 μm (UPLXAPO100XO, Olympus), the amount of PDMS poured over the mold should be strictly controlled within 5-10 mg to guarantee a thickness of the PDMS layer on glass < 50 μm.

Step 2: Nanodiamond coating on PDMS pillar array.

The above PDMS pillar array surface was aminated by emerging into the (3-Aminopropyl)triethoxysilane (APTES) solution in ethanol (10%, wt/v) for 20 min, and then washed by ethanol and DI water, respectively. SAAO nanodiamond sample (0.05 mg/mL in water) was dispersed by ultrasonic wave for 30 min. Incubate the dispersed SAAO nanodiamond sample with the PDMS pillar array for 1.5 h and wash away the unabsorbed NDs. The obtained NDs@pillar array samples were then incubated with fibronectin solution (50 μg/mL) for 4 h. After that, the sample was washed and stored in 4°.

**COMSOL simulation**

The deformation of the PDMS micropillar was simulated with the Solid Mechanics interface in COMSOL Multiphysics® 6.0. Specifically, the micropillar was modelled as a cylinder (2 μm in diameter) with the bottom surface attached to a rigid plate and the top surface in direct contact with the cell. The filopodia traction experienced by the micropillar was represented by a surface load of 50 nN, evenly distributed in a small circular region (with a diameter of 0.2 μm) on the top surface. Young's modulus of the micropillar was fixed at 1.4 MPa, while the aspect ratio, Poisson's ratio, and loading position were varied within certain ranges. Tetrahedron elements were used to discretize the micropillar, with extra mesh refinement implemented within the loading area. The computation was carried out by stationary solver with the relative tolerance setting as $10^{-6}$.

**Cell culture**

NIH-3T3 cells were cultured in DMEM (Gibcol) supplemented with 10% (v/v) FBS (Invitrogen), 2 mM L-glutamine (Invitrogen), and 100 U/mg penicillin-streptomycin (Invitrogen) at 37 °C and 5% $CO_2$. Before cell seeding, diamond sensor array was coated with human plasma fibronectin (50 μg/mL; Thermo) and incubated in 4° for 4 h. The obtained sensor arrays were then washed by phosphate-buffered saline (PBS) 3 times and sterilized by UV light for 20 minutes. Cells were then trypsinized, suspended in DMEM and plated on the sensor 4-6 h. Human monocytic THP-1 cells (American Type Culture Collection) were cultured in suspension in RPMI-1640 medium (Gibco) supplemented with 10% (v/v) heat-inactivated FBS (Invitrogen), 0.05 mM 2-mercaptoethanol (Sigma-Aldrich), and 100 U/mL penicillin-streptomycin (Invitrogen) at 37 °C and 5% $CO_2$. To obtain freshly differentiated macrophage cells, THP-1 monocytes were seeded at a density of $5 \times 10^5$ cells/mL and treated with transforming growth factor beta 1 (TGF-β1, 100 nM; Sigma-Aldrich, T7039) for 24-48 h (*32*). For inhibitor experiments, cells were plated on culture dishes or sensors at a density of 20000 cells $cm^{-2}$ in the presence of following inhibitors: Y27632 (20 μM, MCE) and Latrunculin B (10 μM, MCE). Inhibitors remained in the media for the duration of the experiments.

**Lentiviral-based transfection**

The plasmid of lifeact-EBFP2 was a gift from Pakorn Kanchanawong. After subcloning into the lentiviral vector, the expression plasmid was combined with packaging plasmid (psPAX2; Addgene #12260) and envelope plasmid (pMD2.G; Addgene #12259) in serum-free DMEM medium, followed by the addition of PolyJet transfection reagent (SignaGen Laboratories, SL100688) and incubate at room temperature for 10 min incubation. The mixture was evenly dropwise into 50-80% confluence 293T cells. After replacing medium with complete DMEM after 12 h, virus-rich supernatant was harvested at 72 h post-transfection. The culture medium was collected and centrifuged at 1,500 rpm for 5 minutes. The supernatant was further filtered with 0.45 μm filter to remove cell debris. THP-1 cells were pretreated with polybrene (10 μg/mL; Sigma, TR-1003) for 10 minutes at room temperature to enhance the membrane permeability and then transfected with the filtered viral solution for 72 hours in a 37 °C incubator with 5% $CO_2$. Puromycin was used to select and maintain the transfected cells.

**Immunofluorescence staining**

After allowing the cells to grow on the sensor surfaces for specific time, the cells were fixed with 4% formaldehyde for 20 min at room temperature, permeabilized with 0.25% Triton X-100 in PBS for 10 min, and then blocked with 1% bovine serum albumin in PBST (PBS with 0.1% Triton X-100) for 45 min at room temperature. Next, the cells were incubated with primary antibodies overnight at 4 °C. The following primary antibodies were used: mouse monoclonal anti-Vinculin (Invitrogen 14-9777-82, 1:100 dilution). Cells were washed with PBS two times and then incubated with the appropriate secondary antibodies (Invitrogen, Goat anti-Mouse IgG Alexa Fluor 488 A-11029) for 1 h at room temperature. Finally, the cells were stained with DAPI (Sigma D9542, 1:1000 dilution) and Phalloidin (Abcam, iFluor 647, ab176759 or iFluor 488, ab176753, 1:1000 dilution) and imaged by Zeiss 980 confocal microscope.

**Sample preparation for SEM**

After allowing the cells to grow on the sensor surfaces overnight, the cells were fixed with 4% formaldehyde for 20 min at room temperature. After that the fixed cells were dehydrated by incubation in a series of ethanol solutions with 30% (5 min), 50% (5 min), 70% (5 min), 80% (10 min), 95% (15 min) and 100%. The gradually increased treatment will remove the water without causing specimen shrinkage. The obtained cell samples were further dried by a critical point drier and then coated with Au (10 nm) for SEM imaging (Leo 1530, Zeiss).

**Fitting model of LPM curve**

The fitting model has been reported in our previous work:

$$I_{flu} = A_1 - A_2 cos^2(\alpha - \beta)$$

Where $A_1, A_2, \alpha$ are fitting parameters ($A_1 > 0, A_2 > 0$), and $I_{flu}$ (fluorescence intensity), $\beta$ (polarization direction of excitation laser) are input parameters.

# DTM workflow and analysis

The rotational angle determination is similar to the conventional standard TFM protocol that compares the phage changes before and after the force relaxation process induced by adding cell lysis reagent (Proteinase K, 50 µg/mL and 0.01% SDS for 30 min). Before the LPM test for determining the in-plane orientation of targeted NV centers, a second-order correlation function with delay time of zero $g^2(0)$ will be performed to determine the NV numbers in each diamond. The optimized fitting model of $g^2(\tau)$ is given by the equation:(23, 33)

$$g^2_{Cor}(\tau) = [1 - (1+a)*e^{-\frac{|\tau|}{c}} + b*e^{-\frac{|\tau|}{d}} - [(-\rho^2)]/\rho^2$$

Where $a, b, c, d$ are all fitting parameters. $\rho = \frac{S}{S+B}$ is related to the signal-to-background ratio, which is measured independently in each experimental run. When $g^2(0)<0.5$, the number of emitter is regarded as 1. Only the diamond with single NV center will be used in the further test (ratio ~ 30%). The displacement information can be obtained by the bright field image of pillars through the commonly used PillarTracker plugin in Image J.(34)

Assume all the pillars deformed in elastic regime, the torque $M$ can be estimated by the equation:

$$M = \frac{\pi E r^4}{4(1+v)L}\theta$$

where $E$ is the Young's modulus (Pa), $v$ is the poisson's ratio, $r$ presents the radius of pillar (m), $L$ is the height of the pillar (m), $\theta$ is the rotation angle at the tip of the pillar.

The twisting energy and bending energy can be estimated by the following equations:

$$W_{twisting} = \frac{KE}{4L(1+v)}\theta^2$$

where the shape index $K = \frac{\pi r^4}{2}$, $v$ is the poisson's ratio and $L$ is the height of the pillar (m).

$$W_{bending} = \frac{1}{2}\alpha EI(\Delta x)^2$$

where $I = \frac{\pi r^4}{4}$, $\alpha = 3/L^3$, and the $\Delta x$ indicates the displacement at the tip of the pillar.

**System error analysis**

To ensure measurement accuracy for each sample, the measurement error was determined by tracking the pillar rotations not associated with any cells before and after the lysis reagent treatment. Only if the experiment results exceed the system error can we attribute this to effective rotational movement caused by cell force.

**Statistical analysis**

Group differences were conducted by student's test with GraphPad Prism 8. P-values < 0.05 were considered statistically significant (*p < 0.05, **p < 0.01, ***p < 0.001). Correlation analysis was performed by SPSS Version 11.0.

**Acknowledgement**

Z.Q.C. acknowledges the financial support from the National Natural Science Foundation of China (NSFC) and the Research Grants Council (RGC) of the Hong Kong Joint Research Scheme (Project No. N_HKU750/23), HKU seed fund, and the Shenzhen-Hong Kong-Macau Technology Research Programme (Category C project, no. SGDX20230821091501008). Y.L acknowledges financial supported by the Research Grants Council of Hong Kong under the General Research Fund (Grant No. 17210520) and the National Natural Science Foundation of China (Grant No. 12272332). C.H.Y acknowledges financial supported by the Research Grants Council of Hong Kong under the General Research Fund (Grant No. 17120021 and C7070-22E).

**Author contributions:** Z.Q.C., Y.L., and Y.H. conceived the concept of the DTM. Y.H. synthesized the force sensor, performed the measurement, data analysis and wrote the manuscript. L.Z.W. performed the measurement and wrote the required programs. Z.H. and F.S. performed the simulations under the supervision of Y.L. Y.T.W. performed the cell transfection under the supervision of C.H.Y. L.Y.Z. assisted with the SEM test. L.J.M. performed the AFM test. W.Y.X. and X.H.H. assisted with the figure design. Q.W. and C.H.Y assisted the data analysis and general discussion. Z.Q.C., Y.L. and C.H.Y. supervised the project. All authors discussed the results and participated in writing the manuscript.

**Competing interests:** The authors declare no competing interests.

**Data and materials availability:** All data are available in the main text or the supplementary materials.

## References

1. O. Chaudhuri, J. Cooper-White, P. A. Janmey, D. J. Mooney, V. B. Shenoy, Effects of extracellular matrix viscoelasticity on cellular behaviour. *Nature*. **584**, 535–546 (2020).
2. M. H. Jo, P. Meneses, O. Yang, C. C. Carcamo, S. Pangeni, T. Ha, Determination of single-molecule loading rate during mechanotransduction in cell adhesion. *Science (80-. )*. **383**, 1374–1379 (2024).
3. W. Xie, X. Wei, H. Kang, H. Jiang, Z. Chu, Y. Lin, Y. Hou, Q. Wei, Static and Dynamic: Evolving Biomaterial Mechanical Properties to Control Cellular Mechanotransduction. *Adv. Sci.* **10**, 2204594 (2023).
4. O. Collin, S. Na, F. Chowdhury, M. Hong, M. E. Shin, F. Wang, N. Wang, Self-Organized Podosomes Are Dynamic Mechanosensors. *Curr. Biol.* **18**, 1288–1294 (2008).
5. W. R. Legant, C. K. Choi, J. S. Miller, L. Shao, L. Gao, E. Betzig, C. S. Chen, Multidimensional traction force microscopy reveals out-of-plane rotational moments about focal adhesions. *Proc. Natl. Acad. Sci.* **110**, 881–886 (2013).
6. N. Leijnse, Y. F. Barooji, M. R. Arastoo, S. L. Sønder, B. Verhagen, L. Wullkopf, J. T. Erler, S. Semsey, J. Nylandsted, L. B. Oddershede, A. Doostmohammadi, P. M. Bendix, Filopodia rotate and coil by actively generating twist in their actin shaft. *Nat. Commun.* **13**, 1636 (2022).
7. S. T. Kim, Y. Shin, K. Brazin, R. J. Mallis, Z.-Y. J. Sun, G. Wagner, M. J. Lang, E. L. Reinherz, TCR Mechanobiology: Torques and Tunable Structures Linked to Early T Cell Signaling. *Front. Immunol.* **3**, 1–8 (2012).
8. S. T. Kim, K. Takeuchi, Z.-Y. J. Sun, M. Touma, C. E. Castro, A. Fahmy, M. J. Lang, G. Wagner, E. L. Reinherz, The αβ T Cell Receptor Is an Anisotropic Mechanosensor. *J. Biol. Chem.* **284**, 31028–31037 (2009).
9. J. Rossy, J. M. Laufer, D. F. Legler, Role of Mechanotransduction and Tension in T Cell Function. *Front. Immunol.* **9**, 1–11 (2018).
10. A. Cockerell, L. Wright, A. Dattani, G. Guo, A. Smith, K. Tsaneva-Atanasova, D. M. Richards, Biophysical models of early mammalian embryogenesis. *Stem Cell Reports*. **18**, 26–46 (2023).
11. Z. Chen, Q. Guo, E. Dai, N. Forsch, L. A. Taber, How the embryonic chick brain twists. *J. R. Soc. Interface*. **13**, 20160395 (2016).
12. Y. Liu, K. Galior, V. P.-Y. Ma, K. Salaita, Molecular Tension Probes for Imaging Forces at the Cell Surface. *Acc. Chem. Res.* **50**, 2915–2924 (2017).
13. J. M. Brockman, A. T. Blanchard, V. Pui-Yan, W. D. Derricotte, Y. Zhang, M. E. Fay, W. A. Lam, F. A. Evangelista, A. L. Mattheyses, K. Salaita, Mapping the 3D orientation of piconewton integrin traction forces. *Nat. Methods*. **15**, 115–118 (2018).
14. H. Colin-York, C. Eggeling, M. Fritzsche, Dissection of mechanical force in living cells by super-resolved traction force microscopy. *Nat. Protoc.* **12**, 783–796 (2017).
15. F. Xu, S. Zhang, L. Ma, Y. Hou, J. Li, A. Denisenko, Z. Li, J. Spatz, J. Wrachtrup, H. Lei, Y. Cao, Q. Wei, Z. Chu, Quantum-enhanced diamond molecular tension microscopy for quantifying cellular forces. *Sci. Adv.* **10**, eadi5300 (2024).
16. T. Zhang, G. Pramanik, K. Zhang, M. Gulka, L. Wang, J. Jing, F. Xu, Z. Li, Q. Wei, P. Cigler, Z. Chu, Toward Quantitative Bio-sensing with Nitrogen–Vacancy Center in Diamond. *ACS Sensors*. **6**, 2077–2107 (2021).
17. N. Aslam, H. Zhou, E. K. Urbach, M. J. Turner, R. L. Walsworth, M. D. Lukin, H. Park,

Quantum sensors for biomedical applications. *Nat. Rev. Phys.* (2023), doi:10.1038/s42254-023-00558-3.
18. J. Zhang, H. He, T. Zhang, L. Wang, M. Gupta, J. Jing, Z. Wang, Q. Wang, K. H. Li, K. K. Y. Wong, Z. Chu, Two-Photon Excitation of Silicon-Vacancy Centers in Nanodiamonds for All-Optical Thermometry with a Noise Floor of 6.6 mK·Hz-1/2. *J. Phys. Chem. C*. **127**, 3013–3019 (2023).
19. F. Perona Martínez, A. C. Nusantara, M. Chipaux, S. K. Padamati, R. Schirhagl, Nanodiamond Relaxometry-Based Detection of Free-Radical Species When Produced in Chemical Reactions in Biologically Relevant Conditions. *ACS Sensors*. **5**, 3862–3869 (2020).
20. L. Nie, A. C. Nusantara, V. G. Damle, M. V. Baranov, M. Chipaux, C. Reyes-San-Martin, T. Hamoh, C. P. Epperla, M. Guricova, P. Cigler, G. van den Bogaart, R. Schirhagl, Quantum Sensing of Free Radicals in Primary Human Dendritic Cells. *Nano Lett.* **22**, 1818–1825 (2022).
21. T. Fujisaku, R. Tanabe, S. Onoda, R. Kubota, T. F. Segawa, F. T. K. So, T. Ohshima, I. Hamachi, M. Shirakawa, R. Igarashi, PH Nanosensor Using Electronic Spins in Diamond. *ACS Nano*. **13**, 11726–11732 (2019).
22. T. P. M. Alegre, C. Santori, G. Medeiros-Ribeiro, R. G. Beausoleil, Polarization-selective excitation of nitrogen vacancy centers in diamond. *Phys. Rev. B*. **76**, 165205 (2007).
23. L. Wang, Y. Hou, T. Zhang, X. Wei, Y. Zhou, D. Lei, Q. Wei, Y. Lin, Z. Chu, All-Optical Modulation of Single Defects in Nanodiamonds: Revealing Rotational and Translational Motions in Cell Traction Force Fields. *Nano Lett.* **22**, 7714–7723 (2022).
24. T. Zhang, L. Ma, L. Wang, F. Xu, Q. Wei, W. Wang, Y. Lin, Z. Chu, Scalable Fabrication of Clean NanodiamondsviaSalt-Assisted Air Oxidation: Implications for Sensing and Imaging. *ACS Appl. Nano Mater.* **4**, 9223–9230 (2021).
25. F. Xu, J. Chen, Y. Hou, J. Cheng, T. K. Hui, S.-C. Chen, Z. Chu, Super-Resolution Enabled Widefield Quantum Diamond Microscopy. *ACS Photonics*. **11**, 121–127 (2024).
26. P. Ucla, J. Le Chesnais, H. Ver Hulst, X. Ju, I. Calvente, L. Leconte, J. Salamero, I. Bonnet, C. Monnot, H. Moreau, J. Landoulsi, V. Semetey, S. Coscoy, Quantifying Cell Traction Forces at the Individual Fiber Scale in 3D: A Novel Approach Based on Deformable Photopolymerized Fiber Arrays. *bioRxiv* (2024), pp. 1–26.
27. W. J. Polacheck, C. S. Chen, Measuring cell-generated forces: A guide to the available tools. *Nat. Methods*. **13**, 415–423 (2016).
28. W. J. Polacheck, C. S. Chen, Measuring cell-generated forces: A guide to the available tools. *Nat. Methods*. **13**, 415–423 (2016).
29. M. A. Al-Aghbar, A. K. Jainarayanan, M. L. Dustin, S. R. Roffler, The interplay between membrane topology and mechanical forces in regulating T cell receptor activity. *Commun. Biol.* **5**, 40 (2022).
30. P. J. de Pablo, I. A. T. Schaap, F. C. MacKintosh, C. F. Schmidt, Deformation and Collapse of Microtubules on the Nanometer Scale. *Phys. Rev. Lett.* **91**, 098101 (2003).
31. M. Amer, H. Wolfenson, *Measuring Cellular Traction Forces with Micropillar Arrays* (2023), vol. 2600.
32. Y. Qi, C. Yu, PI(3,4,5)P3-mediated Cdc42 activation regulates macrophage podosome assembly. *Cell. Mol. Life Sci.* **82**, 127 (2025).
33. R. Brouri, A. Beveratos, J.-P. Poizat, P. Grangier, Photon antibunching in the fluorescence of individual color centers in diamond. *Opt. Lett.* **25**, 1294 (2000).
34. X. F. M. M. Xiaochun, PillarTracker. *GitHub* (2018), (available at https://imagej.net/plugins/pillartracker).

# Supplementary Materials for

**Nanodiamond-Enabled Torsion Microscopy Uncovers Multidimensional Cell-Matrix Mechanical Interactions**
Yong Hou *et al.*

Corresponding authors:
Prof. Dr. Cheng-han Yu, Email: chyu1@hku.hk
Prof. Dr. Yuan Lin, Email: ylin@hku.hk
Prof. Dr. Zhiqin Chu, Email: zqchu@eee.hku.hk

**This file includes:**

    Figs. S1 to S11
    References (1 to 4)

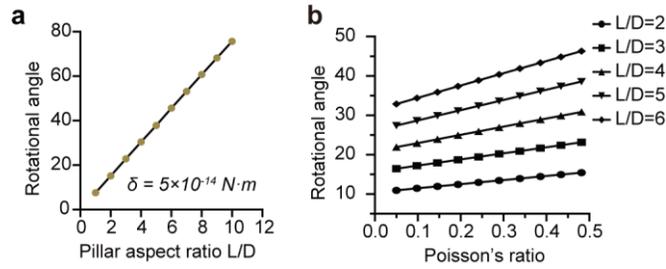

**Figure S1. Computational modeling of how pillar geometry and material properties influence rotational deformation.** a) Modeling the pillar rotational angles as the function of pillar aspect ratio when a torque of $5\times10^{-14}$ N.m is applied on the targeted pillar. b) Modeling the pillar rotational angles as the function of Poisson's ratio and aspect ratio of pillars when a 50 nN force is acting at the edge of pillar with a representative elastic module of 1.4 MPa.

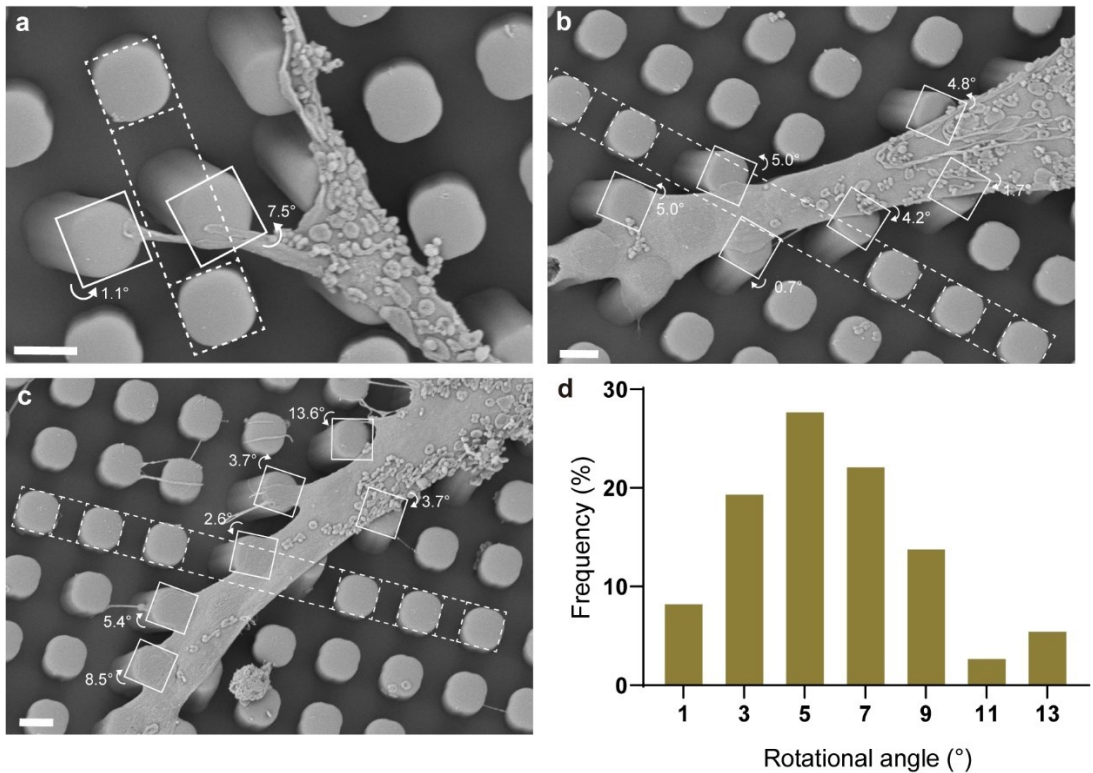

**Figure S2. Square micropillars enable direct visualization of cellular torque.** a-c) SEM images of NIH 3T3 cells spreading on PDMS pillar array with square shape (6 μm in height and 2.5 μm in side length). The pillar edge can be determined by adjusting the image contrast. The rotational angle can be roughly measured by comparing the extension lines of the targeted pillar and reference pillar via Image J. Scale bar: 2 μm. d) Histogram illustrating the distribution of rotational angles of micropillars at the cellular boundary obtained through analysis of SEM images (N = 40).

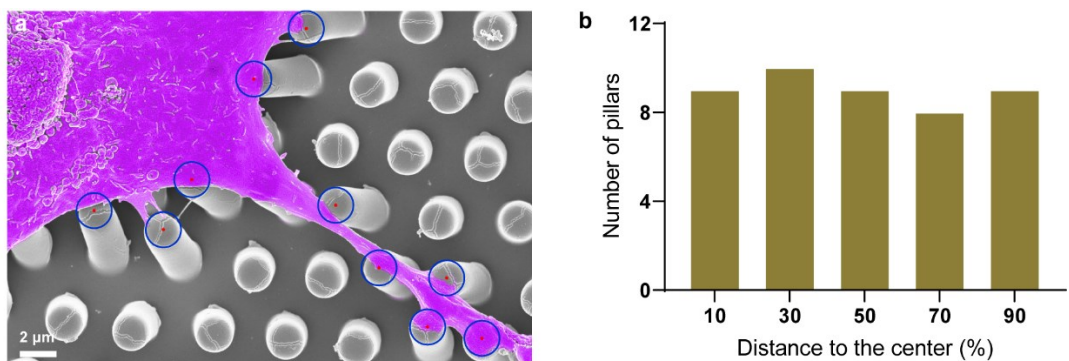

**Figure S3. Off-center force application leads to widespread torsion in cell-pillar interactions.** a) Cell anchoring points at cell boundary are randomly distributed on the pillar top surface. The blue circle indicates the edge of pillar, and the red dots indicate the center of pillars. The cell was labeled with fake color magenta. b) The statistics of force loading position on the pillar (N = 44).

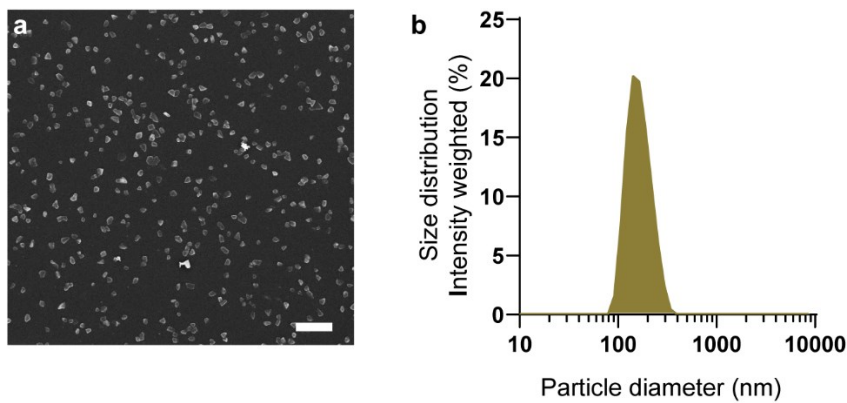

**Figure S4. Characterization of NDs.** a) SEM image of ND. Scale bar: 2 μm; b) Size distribution of NDs.

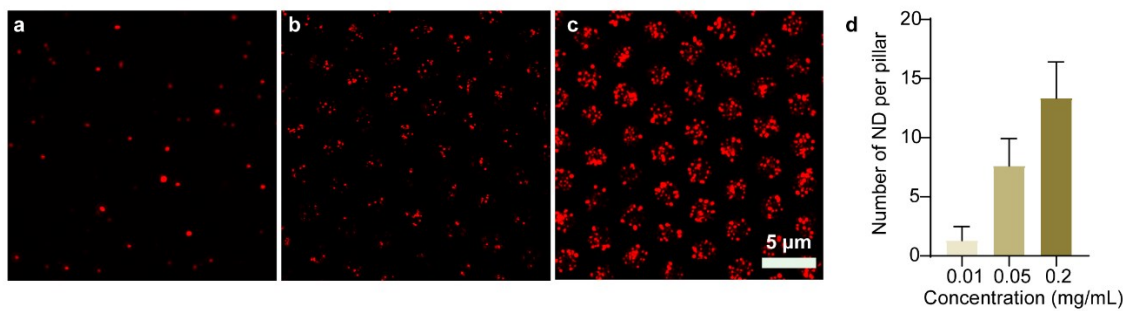

**Figure S5. ND density on pillar tops is tunable via incubation concentration.** The red dots in images show the reflection signals of NDs. a) 0.01 mg/mL; b) 0.05 mg/mL; c) 0.2 mg/mL; d) Statistics of ND density on individual pillars.

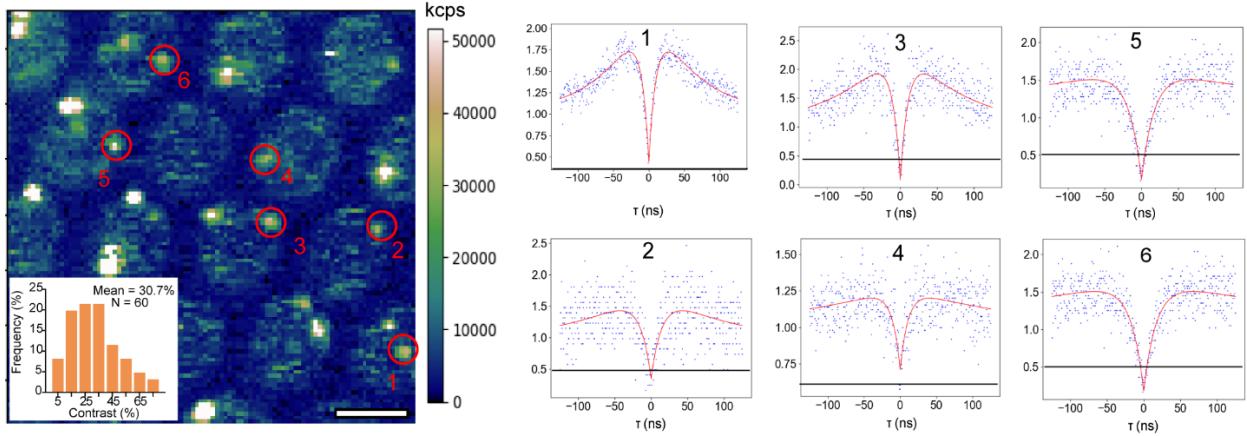

**Figure S6. Photon antibunching confirms single NV centers suitable for LPM measurement.** The measured $g^2(\tau)$ for 6 randomly selected NDs on the pillar array. The result shows the $g^2(0)$ values of ND 2, 5, 3, and 6 are lower than 0.5 that can be used in LPM test. The insert shows the histogram of the contrast of NV center under the polarization modulation. Scale bar: 2 μm.

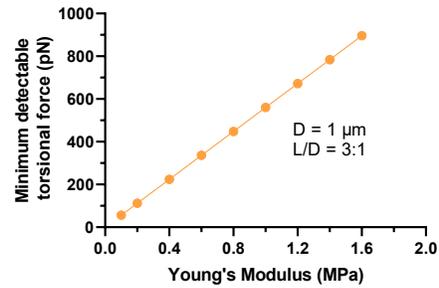

**Figure S7. Torque sensitivity of DTM enhances with pillar mechanical engineering.** Calculation of the minimum detectable torsional force of DTM as the function of Young's modulus, for a pillar diameter of 1 μm and aspect ratio of 3:1. Here, the minimum detectable torsional force is defined as the magnitude of the force required under ideal conditions—applied at the edge of the micropillar top—to induce a rotation of 1.49° (the angular detection limit of DTM).

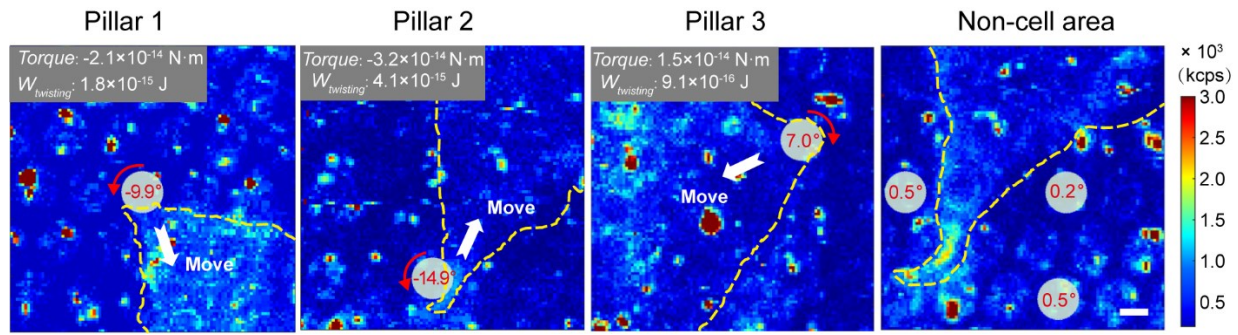

**Figure S8. DTM quantifies local torque and twisting energy across selected pillars.** Fluorescence images of selected areas showing the rotational angles, torque, and twisting energy of the selected pillars. Scale bar: 2 μm.

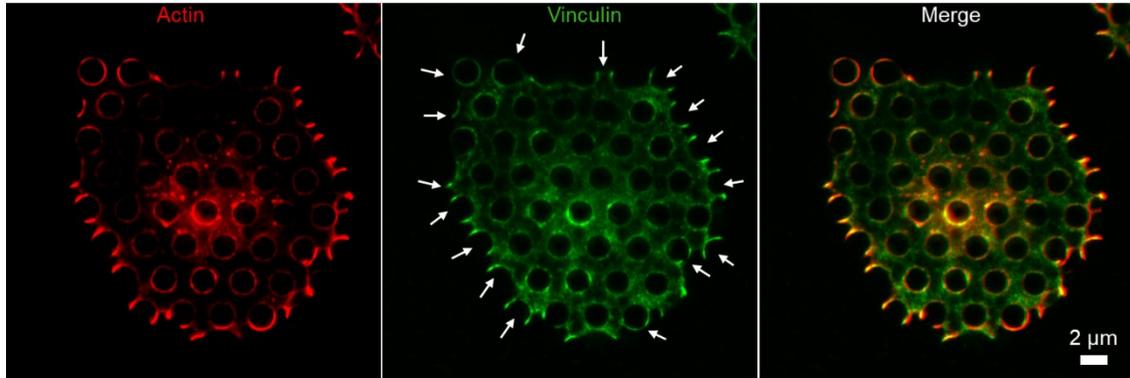

**Figure S9. Curved vinculin structures indicate off-center force transmission.** Fluorescent images of cells spreading on pillar array (culture for 4 h). Actin (red), vinculin (green). The white arrows indicate the curved and circular focal adhesion structures on the pillar.

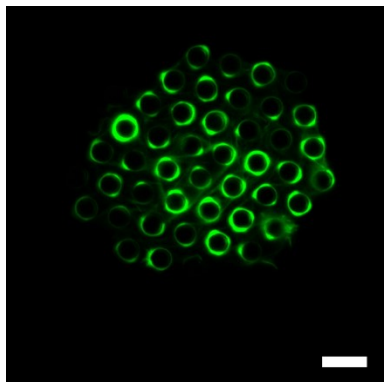

**Figure S10. Macrophages form ring-like adhesions distributed uniformly across the ventral surface.** Fluorescence image of a macrophage cultured on pillars for 24 h in induction medium plus 48 h in growth medium, stained for vinculin (green). Scale bar: 4 μm.

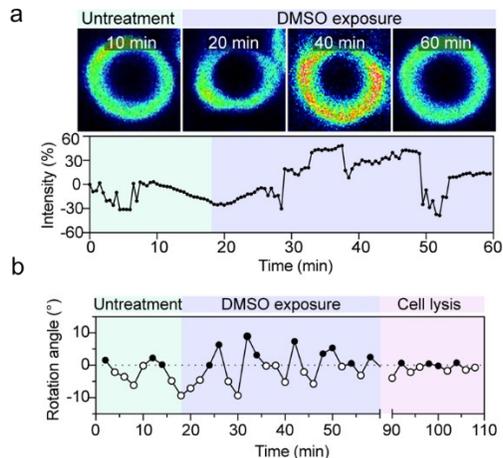

**Figure S11. Live-cell imaging reveals dynamic torsional oscillations in macrophage adhesions.** a) Live fluorescent imaging of ring-like adhesion complex of macrophage on micropillar sensors treated with DMSO (Control group for inhibition test). Fluorescent images were taken at 30s intervals over a 60-minute period. b) Representative profiles of ND rotational angles in macrophages cultured on micropillar substrates with DMSO (Control group for inhibition test). Measurements were conducted at 2-minute intervals over 110 min, with a 30-minute pause for cell lysis treatment. Data represent changes in rotational angle calculated as differences between consecutive time points.

**Reference**
1. Wang, L. *et al.* All-Optical Modulation of Single Defects in Nanodiamonds: Revealing Rotational and Translational Motions in Cell Traction Force Fields. *Nano Lett.* **22**, 7714–7723 (2022).
2. Amer, M. & Wolfenson, H. *Measuring Cellular Traction Forces with Micropillar Arrays*. *Methods in Molecular Biology* vol. 2600 (2023).
3. Brouri, R., Beveratos, A., Poizat, J.-P. & Grangier, P. Photon antibunching in the fluorescence of individual color centers in diamond. *Opt. Lett.* **25**, 1294 (2000).
4. Xiaochun, X. F. M. M. PillarTracker. *GitHub* https://imagej.net/plugins/pillartracker (2018).